# DAG: Deep Adaptive and Generative $K$-Free Community Detection on Attributed Graphs

Chang Liu*
isonomialiu@sjtu.edu.cn
Shanghai Jiao Tong University
Shanghai, China

Yuwen Yang
youngfish@sjtu.edu.cn
Shanghai Jiao Tong University
Shanghai, China

Yue Ding†
dingyue@sjtu.edu.cn
Shanghai Jiao Tong University
Shanghai, China

Hongtao Lu
htlu@sjtu.edu.cn
Shanghai Jiao Tong University
Shanghai, China

Wenqing Lin†
edwlin@tencent.com
Tencent, Shenzhen, China

Ziming Wu
jimmyzmwu@tencent.com
Tencent, Shenzhen, China

Wendong Bi
wendongbi@tencent.com
Tencent, Shenzhen, China

## ABSTRACT
Community detection on attributed graphs with rich semantic and topological information offers great potential for real-world network analysis, especially user matching in online games. Graph Neural Networks (GNNs) have recently enabled Deep Graph Clustering (DGC) methods to learn cluster assignments from semantic and topological information. However, their success depends on the prior knowledge related to the number of communities $K$, which is unrealistic due to the high costs and privacy issues of acquisition. In this paper, we investigate the community detection problem without prior $K$, referred to as $K$-Free Community Detection problem. To address this problem, we propose a novel Deep Adaptive and Generative model (DAG) for community detection without specifying the prior $K$. DAG consists of three key components, *i.e.,* a node representation learning module with masked attribute reconstruction, a community affiliation readout module, and a community number search module with group sparsity. These components enable DAG to convert the process of non-differentiable grid search for the community number, *i.e.,* a discrete hyperparameter in existing DGC methods, into a differentiable learning process. In such a way, DAG can simultaneously perform community detection and community number search end-to-end. To alleviate the cost of acquiring community labels in real-world applications, we design a new metric, EDGE, to evaluate community detection methods even when the labels are not feasible. Extensive offline experiments on five public datasets and a real-world online mobile game dataset demonstrate the superiority of our DAG over the existing state-of-the-art (SOTA) methods. DAG has a relative increase of 7.35% in teams in a Tencent online game compared with the best competitor.

## 1 INTRODUCTION
Community detection, which aims to partition networks into densely connected substructures and reveals latent functions [12, 24], is a crucial unsupervised learning task in network analysis. It has been extensively studied in various fields, such as recommendation systems [23, 44, 62], biochemistry [11, 50, 55], cyber security [54], and business [2, 3, 25]. Among various networks, attributed networks, where nodes contain abundant semantic information, have gained significant attention in recent years since node attributes can play a complementary role of the network topology [35, 40, 53, 57, 61]. Its efficacy is evident that nodes with similar attributes tend to form cohesive communities in real-world social networks, as suggested by the adage "birds of a feather flock together" [37].

Existing algorithms for community detection in attributed networks suffer from two limitations in industrial applications: 1) From a learning perspective, it is not feasible to concurrently acquire representations from network topology and node semantics while also searching for the optimal community number, denoted as $K$. Specifically, conventional community detection algorithms

---

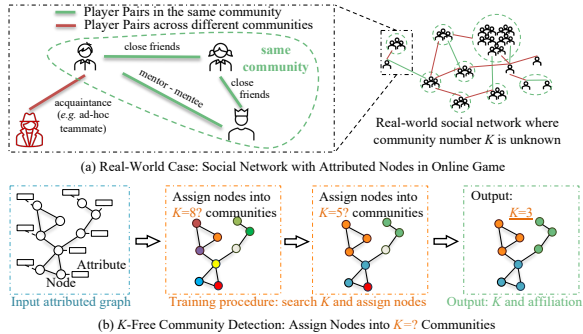

Figure 1: Our research problem: $K$-Free community detection for real-world social networks (node-attributed graphs).

struggle to strike a balance between learning the intricate network topology and handling high-dimensional semantic attributes [49]. In comparison, advanced deep learning-based methods achieve topology-semantic trade-off, but they rely on prior knowledge of $K$, rendering their practical application challenging. 2) From the evaluation perspective, the ground-truth community is a trade-off between topology and semantics, making unsupervised metrics deviate from the real-world communities. Meanwhile, the labels in large-scale social networks are unavailable for user privacy, making existing metrics infeasible in deployment.

To support our investigation, we evaluate traditional methods on five well-known datasets with unsupervised metrics, *i.e.,* modularity [7] and Calinski Harabasz score (dubbed as "Semantic") [6] to measure the node connection density and the attribute similarity in the same community. As illustrated in Table 1, the ground truth community is a *trade-off* between topology and semantics, but existing methods overemphasize a specific metric, *i.e.,* Louvain only focuses on network connections, and attribute clustering algorithms like $K$-means [14] solely concentrate on attributes. This imbalance results in biased outcomes and deviates from the real-world community structure. It should be noted that while Louvain can adaptively search for the community number $K$ ($K$-means cannot), its result is much greater than the ground truth across all datasets. Thus, it falls short to discover a suitable $K$.

Deep Graph Clustering (DGC) [34] methods employ Graph Neural Networks (GNNs) and unify the learning from both topological structure and node semantic attributes by learning "clustering-friendly" node embeddings [38]. However, they fail to address the dependency of knowing $K$, which precludes the applicability of real-world community detection. One straightforward solution is to estimate $K$ by traditional methods, but it is often larger than the ground truth. Thus, naively estimating $K$ via traditional methods and subsequently applying DGC may fall into sub-optimal. DGC circumvents this problem by assuming $K$ is already known, which is unrealistic in practice. An example is illustrated in Fig. 1 (a) for user communities in online games. Ground truth labels can be some private user profiles, such as affiliation, job, location, *etc.*, which are not available to the platform. We can only identify users with frequent interactions and high attribute similarity. These users are likely to belong to the same community, and it is still challenging to determine the number of communities they form. This exhibits a "catch-22" dilemma: existing deep learning approaches necessitate prior knowledge, but it does not exist. Consequently, there is an urgent need to detect communities with unknown community number issues in real-world attributed graphs [32].

| Dataset | Algorithm | Modularity | Semantic | K |
|---|---|---|---|---|
| Cora [46] | Ground Truth | 0.6401 | 11.936 | 7 |
| | $K$-means | 0.1933 | 20.962 | 7 |
| | Louvain | 0.8135 | 2.107 | 105 |
| CiteSeer [46] | Ground Truth | 0.5470 | 11.646 | 6 |
| | $K$-means | 0.2970 | 19.349 | 6 |
| | Louvain | 0.8919 | 1.615 | 469 |
| PubMed [46] | Ground Truth | 0.4318 | 200.337 | 3 |
| | $K$-means | 0.3490 | 435.917 | 3 |
| | Louvain | 0.7695 | 37.069 | 39 |
| Wiki [56] | Ground Truth | 0.5420 | 11.368 | 17 |
| | $K$-means | 0.2061 | 24.986 | 17 |
| | Louvain | 0.7112 | 3.530 | 64 |
| CoraFull [47] | Ground Truth | 0.5417 | 10.468 | 70 |
| | $K$-means | 0.2462 | 22.371 | 70 |
| | Louvain | 0.8126 | 2.344 | 404 |

Table 1: Traditional methods lead to biased results from ground-truth communities. Modularity measures the density of the communities. The "Semantic" metric is the Calinski Harabasz score. $K$ is the community number.

In this paper, we aim to develop a systematic solution for the $K$-free community detection problem, as illustrated in Fig. 1 (b). We design a novel learning framework named **D**eep **A**daptive and **G**enerative (DAG) for community detection on node-attributed graphs. Specifically, DAG first learns node embeddings with both topology and semantic information with masked attribute reconstruction. Secondly, we design a community readout module based on the community affiliation network [5, 26] instead of clustering, which is the key difference between DAG and DGC methods. The readout module enables our third step for differentiable community selection. We convert the challenging grid search of $K$ for clustering into a differentiable community selection regularized by group sparsity. In summary, DAG does not require specifying prior $K$ but simultaneously performs community detection and community number search in an end-to-end fashion. We additionally propose a novel metric, EDGE, to address the high acquisition costs for evaluation. EDGE transforms the $K$ class problem into a binary one to replace the unavailable private profile with high-confidence user interaction in deployment. Empirical experiments justify that EDGE is more robust than existing metrics when the detected $K$ is not always equal to the ground truth in Sec.4.2.1 and can indicate more meaningful communities where users are more likely to interact with each other in Sec. 5.

**Contributions**. The contributions of our paper include:
- We propose a $K$-free deep community detection framework on attributed graphs called DAG, which can adaptively search the number of communities during the training process in an end-to-end manner. DAG bridges the gap between traditional and deep learning-based community detection methods.
- We design a new EDGE metric for $K$-free community detection evaluation. EDGE offers two advantages: 1) For labeled data, EDGE is robust and objective if detected $K$ varies from ground truth. 2) EDGE is also effective for real-world applications in which we do not have actual ground truth communities because EDGE reflects the intimacy between linked nodes.
- We conduct extensive experiments on five public benchmark datasets in Sec. 4. Experimental results demonstrate that DAG outperforms state-of-the-art (SOTA) methods. We further conduct an online A/B testing between our DAG and the best SOTA against the baseline on a friend recommendation task during a one-week event; the results show that DAG's improvement outperforms SOTA by 7.35%, 1.97%, and 5.24% for the overall success rate, click rate, and team formation success rate, respectively. The superior online performance further indicates that DAG can detect more meaningful user communities, *i.e.*, users within the same community have a higher tendency to interact.

## 2 PRELIMINARIES

In this section, we first formulate our research problem, *i.e.*, $K$-free community detection on attribute graphs. We then conduct an empirical study on DGC methods, demonstrating that they can neither handle $K$-free community detection tasks with trivial modifications nor find a proper community number $K$ with their methods.

### 2.1 Problem Formulation

In an undirected and unweighted attributed graph $\mathcal{G} = \{\mathbf{A}, \mathbf{X}\}$, let $\mathcal{V} = \{v_1, v_2, \cdots, v_N\}$ be a set of $N$ nodes and $\mathcal{E}$ be a set of edges. $\mathbf{X} \in \mathbb{R}^{N \times D}$ and $\mathbf{A} \in \mathbb{R}^{N \times N}$ denote the node attribute matrix and original adjacency matrix, respectively. We define community affiliation as follows to represent node-to-community assignment.

DEFINITION 1 (COMMUNITY AFFILIATION). *A community affiliation $\mathbf{C}_i \in \mathbb{R}_{\geq 0}^K$ of node $v_i$ is a stochastic vector that adds up to one, where the $k$-th entry is the probability of node $v$ belonging to the $k$-th community.*

Based on Definition 1, we focus on a non-overlapping community detection task, *i.e.*, each node belongs to only one community. However, unlike existing DGC methods, we do not have prior knowledge of the total number of communities, denoted as $K$. We formulate the $K$-free community detection problem as follows.

PROBLEM 1 ($K$-FREE COMMUNITY DETECTION). *The task of $K$-free community detection involves determining a community number $K$ and a community affiliation matrix $\mathbf{C} \in \mathbb{R}_{\geq 0}^{N \times K}$ for all $N$ nodes in a given attributed graph $\mathcal{G} = \{\mathbf{A}, \mathbf{X}\}$.*

The objective of $K$-free community detection is to ensure that nodes within a community exhibit stronger topological connections and share more common characteristics compared to nodes in different communities, such as external ground truth labels (if available), connectivity patterns, and node features.

### 2.2 Empirical Investigations of DGC Methods

We conduct empirical studies to investigate the impact of the unknown number of communities $K$ for DGC methods.

Firstly, we choose the SOTA model CCGC as the base model. We then replace CCGC's $K$-means clustering by DBSCAN [9], which is a density-based clustering method that does not rely on the prior $K$. Fig. 2 illustrates the results on the Cora dataset. We project node embeddings on the two-dimensional space via T-SNE [51], and we observe that the vanilla CCGC with $K$-means clustering groups embeddings into $K$ clusters. At the same time, CCGC with DBSCAN situates nodes on a manifold without distinguishable gaps. This change poses challenges for DGC methods, as their training procedure relies on clustering results as soft or hard labels. Consequently, the optimization objective of DGC methods varies across epochs, potentially leading to their collapse. Moreover, hyperparameter tuning for DBSCAN becomes more challenging, as fine-tuning the radius parameter and the minimum number of points in DBSCAN is considered difficult [9, 10, 45]. Additionally, each epoch requires adjusting DBSCAN since node embeddings have been modified based on the previous epoch training. We conclude that in real-world scenarios where $K$ is unknown, clustering-based self-supervised learning methods may collapse due to uncertain training objectives. Existing DGC methods are unable to handle $K$-free community detection tasks with trivial modifications.

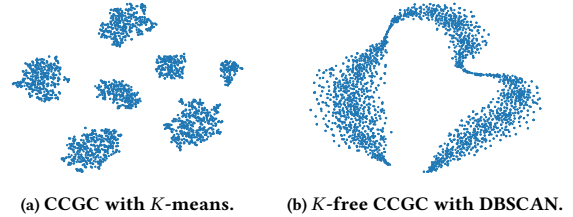

(a) CCGC with $K$-means.    (b) $K$-free CCGC with DBSCAN.

Figure 2: T-SNE [51] visualization of Cora dataset's node representations by deep graph clustering method CCGC [57].

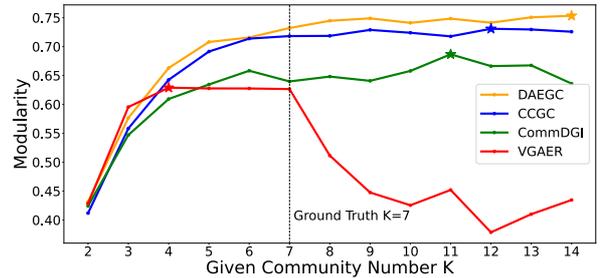

Figure 3: Community number with the highest modularity does not match the ground truth $K$ on Cora.

Next, we traverse the prior $K$ of the DGC methods and observe the changes in the unsupervised modularity metric. This is a searching strategy mentioned by several DGC methods [13, 28, 41] to validate their performance on graphs without known $K$. Our goal is to determine if this strategy can effectively capture real-world community structures. We implement several DGC methods, namely DAEGC [53], CommDGI [61], VGAER[41], and CCGC [57] as examples. For the Cora dataset [46], we observe that their estimated community number with the highest modularity does not align with the ground truth $K$. This discrepancy persists even when considering a range of $[2, 2 \times K_{\text{GT}}]$, where $K_{\text{GT}}$ is the ground truth community number. Furthermore, in real-world applications such as online games, online tests are often employed to collect user activity as the final metric. Online tests typically require days to gather reliable results; real-world graphs are large-scale, and user communities may change over time, rendering grid search on $K$ impractical. Therefore, selecting the number $K$ of communities with the highest unsupervised modularity index through traversal and repeated training is neither efficient nor effective. It becomes a time-consuming process that does not guarantee optimal results.

From these observations, we conclude the findings that:
(1) In real-life scenarios where $K$ is unknown, clustering-based self-supervised learning methods can collapse due to uncertain training objectives.
(2) It is neither efficient nor effective to select the number of communities $K$ with the highest unsupervised modularity through traversal and repeated training.

In the following sections, we propose our DAG method to tackle this challenging task and overcome the aforementioned issues.

## 3 DEEP ADAPTIVE AND GENERATIVE COMMUNITY DETECTION

In this section, we aim to address two challenges of $K$-free community detection, *i.e.,* how to detect communities without $K$, and evaluate the results in a low-cost and robust manner. We propose a general framework, named DAG, to jointly learn node embedding $\mathbf{H}$, community affiliation $\mathbf{C}$, and community number $K$. As shown in Fig. 4, the key insight of DAG is introducing a Community Affiliation Network (CAN) based generative model instead of clustering-based DGC methods, converting the non-differentiable $K$ searching problem to a differentiable one and solving it with group sparsity. We also design an EDGE metric to convert a $K$ class problem into a binary edge classification problem.

### 3.1 Masked Attribute Reconstruction

In attributed graph community detection, obtaining node representations that incorporate structural and semantic aspects is crucial. To achieve this, inspired by recent progress in masked auto-encoders for node classification [20, 21], we introduce the Masked Attribute Reconstruction module, which is trained with a task that randomly masks attributes and reconstructs them. This process encourages the node representation to incorporate both its attributes and the attributes of its topological neighbors.

For the graph $\mathcal{G} = (\mathbf{A}, \mathbf{X})$ with node set $\mathcal{V}$, we sample a set of nodes $\widetilde{\mathcal{V}} \subset \mathcal{V}$ for each epoch, and replace their attribute vectors with a learnable [Mask] Token $\mathbf{X}_{[M]} \in \mathbb{R}^D$:

$$\widetilde{\mathbf{X}} = \text{MASK}(\mathbf{X}), \quad \text{where } \widetilde{\mathbf{X}}_i = \begin{cases} \mathbf{X}_{[M]} & v_i \in \widetilde{\mathcal{V}} \\ \mathbf{X}_i & v_i \notin \widetilde{\mathcal{V}} \end{cases}. \quad (1)$$

We use two layers of GAT [52] as the encoder to encode the masked graph and generate the node representation matrix $\mathbf{H} \in \mathbb{R}^{N \times D'}$, where $D'$ is the embedding length of each node:

$$\mathbf{H} = \text{Encoder}(\mathbf{A}, \widetilde{\mathbf{X}}). \quad (2)$$

This embedding $\mathbf{H}$ will simultaneously perform two tasks: attribute reconstruction and community detection. The ReMask trick is employed to encourage the model's embedding further to contain

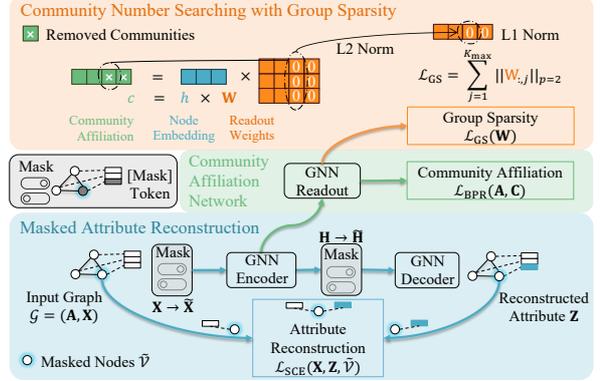

Figure 4: DAG framework.

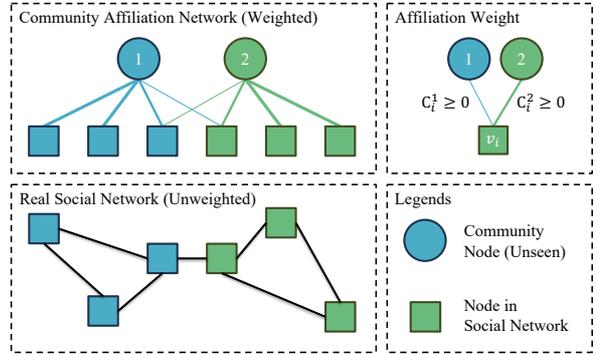

Figure 5: The Community Affiliation Network.

semantic information about its topological neighborhood:

$$\widetilde{\mathbf{H}} = \text{MASK}(\mathbf{H}). \quad (3)$$

We use two GAT layers as attribute Decoder to output the restored feature matrix:

$$\mathbf{Z} = \text{Decoder}(\mathbf{A}, \widetilde{\mathbf{H}}). \quad (4)$$

We introduce the Scaled Cosine Error (SCE) proposed by Graph-MAE [21]. This is because the feature vectors of attribute graphs are often very sparse, and the MSE loss easily converges to a trivial solution of all zeros.

$$\mathcal{L}_{\text{SCE}}(\mathbf{X}, \mathbf{Z}, \widetilde{\mathcal{V}}) = \frac{1}{|\widetilde{\mathcal{V}}|} \sum_{v_i \in \widetilde{\mathcal{V}}} \left(1 - \frac{x_i^T z_i}{\|x_i\| \cdot \|z_i\|}\right)^3. \quad (5)$$

In summary, Masked Attribute Reconstruction learns node representations combining topological and semantic aspects, which is essential for the subsequent community affiliation readout.

### 3.2 Community Affiliation Readout

DAG simultaneously learns node representations and community affiliations to enable further searching $K$ end-to-end. Inspired by Community Affiliation Network (CAN), a classical social model [5, 26, 64] explaining how social networks are generated, we design a readout module to model nodes' affiliation explicitly.

As shown in Fig. 5, CAN is a weighted bipartite graph containing all real nodes and unseen community nodes of a social network. Each real node in the social network is associated with a community affiliation vector $\mathbf{C}_i \in \mathbb{R}^K$, where $K$ is the number of communities, and each entry $\mathbf{C}_{i,j}$ represents the affiliation strength of node $v_i$ to the community $j$. The CAN model reconstructs the graph's adjacency matrix based on the community affiliations, providing a differentiable objective for learning the community structure. When reconstructing the adjacency matrix, the similarity of the community affiliation vectors $(\mathbf{C}_i, \mathbf{C}_j)$ of a node pair $(i, j)$ indicates the probability of generating an edge $p(i, j)$ between them:

$$p(i, j) = \sigma(\mathbf{C}_i \cdot \mathbf{C}_j^T), \qquad (6)$$

where $\sigma$ is the sigmoid [36] function.

The input embedding matrix $\mathbf{H}$ is augmented throughout training and can contain information about its neighbor nodes. Based on this property, we use a two-layer GCN and a softmax [17] function as the community affiliation readout to output the community affiliation $\mathbf{C} \in \mathbb{R}^{K_{\max}}_{\geq 0}$ of all nodes, where $K_{\max}$ is the maximum possible community number.

$$\mathbf{C} = \text{Readout}(\mathbf{A}, \mathbf{H}). \qquad (7)$$

Since we do not have the actual community number $K$ in the $K$-free community detection scenario, we set $K_{\max}$ to a relatively large number. The $K_{\max}$ setting can be found in Sec. 4.

The community ID $I_i$ of node $i$ is the number of digits where the maximum value of the Community matrix exists, which is similar to node classification:

$$I_i = \arg\max_j \mathbf{C}_{i,j}. \qquad (8)$$

Generating the whole adjacency matrix of the network requires a complexity of $O(n^2)$, which is not realistic for large graphs. Therefore, Bayesian Personalized Ranking (BPR) loss [43] is used to predict each existing edge $(i, j)$ by sampling a negative edge $(i, u)$:

$$\mathcal{L}_{\text{BPR}} = -\sum_{i=1}^{|\mathcal{E}|} \ln \sigma(\mathbf{C}_i \cdot \mathbf{C}_j^T - \mathbf{C}_i \cdot \mathbf{C}_u^T), \qquad (9)$$

where $|\mathcal{E}|$ is the number of edges.

In summary, the Community Detection Readout module, which is the main difference between DAG and DGC methods, simultaneously learns node representations and community affiliations in an end-to-end manner. This end-to-end readout enables the differentiable searching process, making it a crucial component for $K$-free community detection in attributed graphs.

### 3.3 Community Number Search

In DGC methods, determining the optimal number of communities $K$ poses a significant challenge, as it is often used as a hyperparameter for $K$-means-like clustering. This makes it difficult to optimize within the DGC framework. Inspired by traditional community detection methods such as Louvain [4], we propose a differentiable Community Number Search method that adaptively finds the best $K$ by gradually merging smaller communities. Our approach is performed on the output layer of the Community Affiliation Readout module during end-to-end training. This method employs group sparsity constraints to gradually compress the number of communities, merging communities with close links and similar attributes. As a result, our approach enables simultaneous learning of node representations, community affiliations, and community numbers.

The input of the last layer of Community Readout is denoted as $\mathbf{H}_\mathbf{C}$, and the calculation of the Community Matrix $\mathbf{C}$ is given by:

$$\mathbf{C} = \text{ReLU}\left(\hat{\mathbf{A}}\mathbf{H}_\mathbf{C}\mathbf{W}\right), \qquad (10)$$

where $\hat{\mathbf{A}}$ is the normalized adjacency matrix obtained by adding self-loops and row-normalizing the original adjacency matrix $\mathbf{A}$. The matrix $\mathbf{W}$ has dimensions $d \times k_{\max}$, where $d$ is the length of the input embedding, i.e., column number of $\mathbf{H}_\mathbf{C}$. The ReLU activation function is applied element-wise to the matrix product $\hat{\mathbf{A}}\mathbf{H}_\mathbf{C}\mathbf{W}$. The group sparsity constraint based on $\mathcal{L}_{2,1}$ norm is defined as:

$$\mathcal{L}_{\text{GS}} = \|\mathbf{W}^T\|_{2,1} = \sum_{j=1}^{k_{\max}} \|\mathbf{W}_{:,j}\|_{p=2} = \sum_{j=1}^{k_{\max}} \left(\sum_{i=1}^{d} \mathbf{w}_{i,j}^2\right)^{1/2}. \qquad (11)$$

LEMMA 1 (GROUP SPARSITY). *The columns of Community Matrix $\mathbf{C}$ are sparse, i.e., some of its column vectors should be zero vectors.*

From the lemma above, we know that $\mathcal{L}_{\text{GS}}$ constraint has two main benefits: (1) it makes $\mathbf{C}$ more sparse, improving the confidence of community readout, and (2) it concentrates the output on columns of $\mathbf{C}$, allowing for an adaptive number of communities. Additionally, this constraint only affects the parameters of the last layer without influencing the generation of embeddings. The proof of this lemma can be found in Appendix A.

For the community ID vector $I \in \mathbb{Z}^N$ for all $N$ nodes in the graph, our group sparsity ensures that the output range will shrink from $[1, K_{\max}]$ to a smaller range. In other words, the output is the number of communities to which at least one node belongs:

$$K = |\{i : \exists v \in \mathcal{V}, I(v) = i\}|. \qquad (12)$$

In conclusion, the proposed group sparsity method adapts the number of communities during end-to-end training, addressing the challenge of searching $K$ for $K$-free community detection.

**Optimization objective.** The final total training loss is:

$$\mathcal{L} = \mathcal{L}_{\text{SCE}} + \alpha \mathcal{L}_{\text{BPR}} + \beta \mathcal{L}_{\text{GS}}, \qquad (13)$$

where the $\alpha$ and $\beta$ are manually set hyperparameters. Empirically, we find that $\alpha$ and $\beta$ are stable across several public datasets. In other words, we don't need to fine-tune them when it comes to a new dataset. Please refer to Appendix C for more details.

### 3.4 EDGE Metric

There are two-fold challenges when evaluating $K$-Free community detection methods. First, accurate community labels for real-world social networks are unavailable. Second, if the number of detected communities $K$ differs from the ground truth $K_{\text{GT}}$, many existing metrics (e.g. F1 score and accuracy) are infeasible even if we know the ground truth labels for public datasets. To address this, we propose a supervised edge metric suitable for partially known real-world networks and public datasets with ground truth labels. This metric converts the community detection problem into a binary classification problem of whether to cut an edge off.

We treat the edges as inter-community edges and intra-community edges. Let $\mathcal{E}^*_{\text{inter}}$ denote the set of inter-community edges, and $\mathcal{E}^*_{\text{intra}}$ denotes the set of intra-community edges. We can obtain all edge labels for public cases based on node community labels.

$$\begin{aligned}\mathcal{E}^*_{\text{inter}} &= \{(i,j) | \mathbf{A}_{i,j} = 1, I^*(i) \neq I^*(j)\}, \\ \mathcal{E}^*_{\text{intra}} &= \{(i,j) | \mathbf{A}_{i,j} = 1, I^*(i) = I^*(j)\},\end{aligned} \quad (14)$$

where $I^*(i)$ denotes the ground truth label of $i$-th node. After community detection, we generate the set predicted $\mathcal{E}_{\text{inter}}$ and $\mathcal{E}_{\text{intra}}$ with output $I(i)$ like Eq. (14). To balance the binary task, we compute the F1 score where $\mathcal{E}^*_{\text{intra}}$ is the positive set. In other words, the EDGE metric measures whether connected node pairs that belong to the same community can be placed in the same detected community, while node pairs that do not belong to the same community can be placed in different detected communities:

$$\text{EDGE} = 2 \cdot \frac{\frac{|\mathcal{E}^*_{\text{intra}} \cap \mathcal{E}_{\text{intra}}|}{|\mathcal{E}^*_{\text{intra}} \cap \mathcal{E}_{\text{intra}}|+|\mathcal{E}^*_{\text{inter}} \cap \mathcal{E}_{\text{intra}}|} \cdot \frac{|\mathcal{E}^*_{\text{intra}} \cap \mathcal{E}_{\text{intra}}|}{|\mathcal{E}^*_{\text{intra}} \cap \mathcal{E}_{\text{intra}}|+|\mathcal{E}^*_{\text{intra}} \cap \mathcal{E}_{\text{inter}}|}}{\frac{|\mathcal{E}^*_{\text{intra}} \cap \mathcal{E}_{\text{intra}}|}{|\mathcal{E}^*_{\text{intra}} \cap \mathcal{E}_{\text{intra}}|+|\mathcal{E}^*_{\text{inter}} \cap \mathcal{E}_{\text{intra}}|} + \frac{|\mathcal{E}^*_{\text{intra}} \cap \mathcal{E}_{\text{intra}}|}{|\mathcal{E}^*_{\text{intra}} \cap \mathcal{E}_{\text{intra}}|+|\mathcal{E}^*_{\text{intra}} \cap \mathcal{E}_{\text{inter}}|}}. \quad (15)$$

In real-world scenarios, we can consider node pairs with high-confidence interaction (e.g., friends with the highest intimacy or mentor-mentee relationships) as intra-community edges and no historical interaction as inter-community edges.

The EDGE metric effectively evaluates community detection methods on public datasets with known ground truth labels and real-world networks with partially known edge information. This provides a practical approach to assess community detection algorithm performance in real-world scenarios where ground truth community labels are difficult to obtain. Additionally, we find that the widely used NMI is sensitive to the $K$ detected and can overestimate the performance of trivial results; please refer to Sec. 4.2.1.

## 4 EXPERIMENTS ON PUBLIC DATASETS

### 4.1 Experimental Settings

To ensure the fairness and validity of our experimental setup, we highlight several key aspects of our experiments.

**Datasets.** We evaluate DAG on five public datasets (Cora [46], CiteSeer [46], PubMed [46], Wiki [56], and CoraFull [47]). The number of nodes (#Node), edges (#Edge), feature dimensions (#Features), communities (#Comm., if available), and inter-community edges (#Cut) are shown in Table 2.

**Compared methods.** We compare our DAG with four traditional algorithms, *i.e.,* Greedy Q [8], Louvain [4], LPA [42], and Hanp [27]. We also implement five SOTA DGC methods, *i.e.,* DAEGC [53], CommDGI [61], AGCN [40], HSAN [35], and CCGC [57].

**Training procedure.** For each epoch, we sample 50% (75% for PubMed) of the nodes in the dataset for masking and recovering the masked node features. Every 50 epochs, we evaluate the communities detected by DAG using unsupervised metrics. As we aim to address the community detection in an **unsupervised manner**, we select the checkpoint with the highest product of two unsupervised metrics, *i.e.,* modularity and Calinski Harabasz score as the final result, and calculate its NMI and EDGE Metric. More detailed settings can be found in Appendix B.

|          | #Nodes  | #Edges    | #Features | #Comm.  | #Cuts           |
|----------|---------|-----------|-----------|---------|-----------------|
| Cora     | 2,708   | 5,278     | 1,433     | 7       | 1,011           |
| CiteSeer | 3,327   | 4,552     | 3,703     | 6       | 1,212           |
| PubMed   | 19,717  | 44,324    | 500       | 3       | 8,760           |
| Wiki     | 2,405   | 8,261     | 4,973     | 17      | 2,590           |
| CoraFull | 19,793  | 63,421    | 8,710     | 70      | 28,023          |
| GAME     | 209,794 | 2,874,396 | 85        | Unknown | Partially Known |

Table 2: Dataset summary.

**Fair comparison.** One significant difference between DAG and DGC methods is that DAG does not require specifying the number of communities $K$ as a priori, while DGC methods do. The research question we aim to address is how to perform community detection without prior knowledge of $K$. Therefore, for a fair comparison, we adopt the same strategy for finding $K$ for both DAG and DGC methods. Specifically, we choose the value of $K$ that maximizes the product of the unsupervised modularity and the Calinski Harabasz score. This serves as a straightforward approach to balance the topological and semantic similarity of the communities.

Furthermore, to ensure a fair comparison, we set the search range for the number of communities to $[2, 2 \times K_{\text{GT}}]$ for all deep learning-based methods, where $K_{\text{GT}}$ is the ground truth number of communities in each dataset, except for CoraFull. Although our DAG method can search within the range of $[2, 2 \times K_{\text{GT}}]$ for CoraFull, the DGC methods require iterating over all possible values of $K$, making it impractical to search over such a large range ($K_{\text{GT}} = 70$) for CoraFull. Therefore, for CoraFull, we set the search range for both DAG and the DAG methods to $[K_{\text{GT}} - 10, K_{\text{GT}} + 10]$.

**Metric.** Following the SOTA DGC methods [35, 40, 53, 57, 61], we evaluate methods with Normalized Mutual Information (NMI) [48] and our proposed EDGE metric. As mentioned earlier, prior knowledge of the number of communities $K$, which is often difficult to obtain in real scenarios such as user communities in GAME, is evaluated using the EDGE metric of DAG and the best SOTA method. We perform an online test for recommendation tasks during a one-week game event. In the online test setting, we test how the detected communities help to encourage user interactions.

### 4.2 Results in Public Datasets

*4.2.1 Main Results.* We analyze the performance of our proposed DAG method in comparison with traditional community detection algorithms and state-of-the-art DGC methods in Table 3. DAG outperforms all DGC methods in terms of both NMI and EDGE metrics across all public datasets, demonstrating its effectiveness in handling the unknown community detection problem. CCGC is a strong baseline for comparison in real-world scenario experiments. Please refer to Appendix D for more detailed results, including standard deviation and unsupervised metrics.

The EDGE metric, introduced in this paper, proves to be a robust evaluation measure for varying community numbers. Note that when the community number is set equal to the node number (the trivial NULL case), the existing NMI metric tends to overestimate the performance, even achieving the best NMI in CoraFull. However, the EDGE metric does not suffer from this issue, as it assigns a value of 0 to all trivial NULL cases, effectively differentiating between meaningful community structures and trivial cases.

| Dataset | | Cora (K=7) | | | CiteSeer (K=6) | | | PubMed (K=3) | | | Wiki (K=17) | | | CoraFull (K=70) | | |
|---|---|---|---|---|---|---|---|---|---|---|---|---|---|---|---|---|
| Metric | | NMI | EDGE | K | NMI | EDGE | K | NMI | EDGE | K | NMI | EDGE | K | NMI | EDGE | K |
| Traditional CD | Greedy Q | 0.4673 | 0.8864 | 106 | 0.3378 | 0.8395 | 488 | 0.2217 | 0.8584 | 114 | 0.4358 | 0.8478 | 90 | 0.4075 | 0.7290 | 499 |
| | Louvain | 0.4468 | 0.8787 | 104 | 0.3243 | 0.8321 | 469 | 0.2062 | 0.8359 | 39 | 0.4559 | 0.8583 | 64 | 0.4792 | 0.7300 | 404 |
| | Hanp | 0.4010 | 0.7508 | 553 | 0.3402 | 0.7393 | 508 | 0.1770 | 0.7126 | 2037 | **0.4995** | 0.7135 | 885 | 0.5560 | 0.6670 | 2113 |
| | LPA | 0.4142 | 0.7871 | 481 | 0.3377 | 0.7530 | 959 | 0.1804 | 0.7329 | 1924 | 0.4858 | 0.8410 | 396 | 0.5664 | 0.6705 | 2328 |
| Trivial | NULL | 0.3762 | 0 | 2708 | 0.3555 | 0 | 3327 | 0.1937 | 0 | 19717 | 0.4846 | 0 | 2405 | **0.5763** | 0 | 19793 |
| DGC | DAEGC | 0.4587 | 0.8714 | 10.0 | 0.2907 | 0.8302 | 11.5 | 0.1784 | 0.8422 | 4.1 | 0.2200 | 0.7235 | 25.0 | 0.4503 | 0.6882 | 60.1 |
| | CommDGI | 0.3192 | 0.8564 | 9.4 | 0.2911 | 0.8269 | 11.1 | 0.1892 | 0.8496 | 4.9 | 0.1839 | 0.7373 | 31.1 | 0.4467 | 0.6756 | 60.3 |
| | AGCN | 0.2172 | 0.8297 | 10.6 | 0.3160 | 0.8316 | 8.2 | 0.2275 | 0.8504 | 3.8 | 0.1962 | 0.7431 | 22.6 | 0.4721 | 0.6955 | 74.2 |
| | HSAN | 0.4497 | 0.8775 | 4.8 | 0.3128 | 0.8413 | 5.1 | OOM | OOM | OOM | 0.4131 | 0.8375 | 29.5 | OOM | OOM | OOM |
| | CCGC | 0.5051 | 0.8887 | 8.5 | 0.4090 | 0.8447 | 11.9 | 0.1922 | 0.8520 | 4.1 | 0.4079 | 0.8467 | 21.8 | 0.4898 | 0.7047 | 73.6 |
| Ours | DAG | **0.5171** | **0.9004** | 7.4 | **0.4118** | **0.8677** | 6.4 | **0.2828** | **0.8938** | 3.4 | 0.4320 | **0.8629** | 15.7 | 0.4932 | **0.7311** | 68.4 |

Table 3: Average result of supervised metrics and community number on public datasets. Trivial NULL is the case where every single node is treated as a community (i.e., $K = N$). OOM means Out-of-Memory error. Underline shows the best DGC performance. Bold is the best performance for all methods.

| Case | Mask | Sparsity | Cora | CiteSeer | PubMed | Wiki | CoraFull |
|---|---|---|---|---|---|---|---|
| 1 | | | 0.8649 | 0.8182 | 0.8092 | 0.8386 | 0.7002 |
| 2 | ✓ | | 0.8966 | 0.8541 | 0.8735 | 0.8584 | 0.7193 |
| 3 | | ✓ | 0.8803 | 0.8374 | 0.8652 | 0.8468 | 0.7256 |
| DAG | ✓ | ✓ | 0.9004 | 0.8677 | 0.8938 | 0.8629 | 0.7311 |

Table 4: Average EDGE metric of mask attribute generation (Mask) and group sparsity (Sparsity) on public datasets as ablation studies.

| Methods | EDGE | Click Rate | Team Rate | Success Rate |
|---|---|---|---|---|
| Baseline | N/A | 2.59% | 76.72% | 2.00% |
| CCGC | 0.82 | 2.66% (+2.93%) | 76.08% (-0.83%) | 2.03% (+2.08%) |
| Ours | 0.88 | 2.72% (+4.90%) | 80.10% (+4.41%) | 2.18% (+9.44%) |

Table 5: Result on the real-world GAME graph. The relative changes compared to the Baseline are in parentheses.

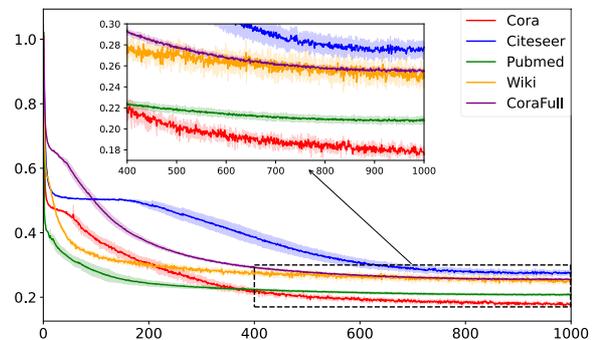

Figure 6: The convergence analysis across different datasets.

In summary, Table 3 shows that our proposed DAG method can effectively handle the K-free community detection problem and outperform SOTA DGC methods, with the EDGE metric serving as a robust evaluation measure.

*4.2.2 Ablation Study.* We conduct an ablation study to investigate the individual contributions of the main components in the DAG model, focusing on the average EDGE metric across the five public datasets. We consider four cases: **Case 1**: Neither mask attribute generation (Mask) nor group sparsity (Sparsity) is applied. **Case 2**: Only mask attribute generation (Mask) is applied. **Case 3**: Only group sparsity (Sparsity) is applied. **DAG**: Both mask attribute generation (Mask) and group sparsity (Sparsity) are applied.

The results in Table 4 show that the DAG model's performance improves with each component. Introducing mask attribute generation (Case 2) improves the EDGE metric, indicating its importance in capturing node semantic information. Applying group sparsity (Case 3) also results in better performance, highlighting its role in adaptively searching for the optimal number of communities.

The full DAG model achieves the highest EDGE metric values across all datasets, demonstrating the effectiveness of combining these components in addressing the K-free community detection problem. In summary, the ablation study confirms the importance of both mask attribute generation and group sparsity components in the DAG model, as their combination leads to the best performance in terms of the EDGE metric across all public datasets.

*4.2.3 Convergence Analysis.* We find that our DAG method converge well across different datasets and hyper-parameter settings. To further show the convergence of DAG, we conduct the following experiments.

**Convergence on different datasets.** We have conducted additional experiments to analyze convergence. We run our DAG on all public datasets with 5 trials. We train our DAG models with 1000 epochs for each trail and record the loss for each epoch without cherry-picking. Finally, we compute the average value (mean) and standard deviation (std) per epoch among the 5 trials. As shown in Fig. 6, we plot the average loss in a line and fulfill the [mean - std, mean + std] with shadow. To make the standard deviations clear, we also zoomed figure that only includes the loss distribution for the last 600 epochs. The results demonstrates that DAG model converges well for different public datasets.

**Convergence on different datasets.** To further draw the concern about DAG's convergence, we also provide the Cora dataset's convergence curve for different scales of our proposed hyper-parameters, i.e., $\alpha$ and $\beta$. Fig. 7 shows that we tune the two hyper-parameters in a log scale. The figure demonstrates that DAG's convergence is stable with different hyper-parameter settings.

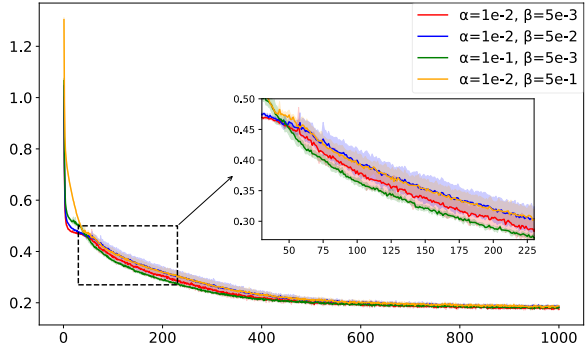

Figure 7: The convergence analysis across different hyper-parameter settings on Cora dataset.

*4.2.4 Case Study.* We provide a case study on the Cora dataset to offer insight into how DAG solves the challenging $K$-free community detection problem. Fig. 8 and Fig. 9 show the community distribution of DAG before and after applying group sparsity. Both models result in relatively uniformly sized communities. However, group sparsity helps DAG merge the detected communities and output some empty communities to search $K$ in an end-to-end manner. Fig. 8 also illustrates that most communities have high purity with respect to the ground truth; however, the large ground truth communities are spliced into several detected communities. For example, the label id 3 is the dominant id in community id=7, 11, and 13. As shown in Fig. 9, with the help of group sparsity, the communities are merged into fewer communities, and DAG merges communities with densely connected and semantically similar communities. As a result, most of the detected communities can better fit the ground truth community.

However, as we can see, the small ground truth community (label id = 6) is not detected by DAG without group sparsity. Consequently, group sparsity fails to merge into a larger community. This also shows that DAG has the potential to improve with a deeper understanding of real-world communities. In summary, DAG demonstrates good performance in solving the $K$-free community detection problem and can be further enhanced with deeper insights into real-world community structures.

## 5 DEPLOYMENT

**Dataset.** Since we want to tackle the $K$-free community detection problem in real-world applications, we evaluate DAG and the best SOTA method on a Tencent mobile *massively multiplayer online role-playing game (MMORPG)* dataset [22, 29–31, 58–60, 62], referred to as GAME. The statistics of GAME can be found in Table 2. We construct the GAME dataset as follows: *(i)* each *daily active user (DAU)* in the game is represented as a node, with the in-game features as attributes, such as the preference for each gameplay style in the game; *(ii)* an edge between two nodes indicates that the two users have friendly relationships, such as friends, mentors, and mentees, among others. We transform this graph into an undirected, unweighted, and homogeneous form to enable a fair comparison using the EDGE metric.

**Competitors and parameter settings.** We select CCGC as the best SOTA DGC method for comparison with our DAG. For this comparison, we directly use the searched $K$ value of DAG to train CCGC. The maximum community number $K_{\max}$ is set to 256. We provide both offline and online experimental results. The best hyperparameters of DAG and CCGC from the public CoraFull dataset are used, as its scale is most similar to the GAME dataset. The detailed parameters can be found in Appendix B.

**Offline experiments.** For offline experiments, we use mentor-mentee relationships as positive examples in the friend network and friends with no intimacy value (*i.e.*, no historical interactions) as negative examples. A higher EDGE score in offline metrics indicates that the algorithm assigns more intimate friends to the same community and less intimate friends to different communities. It is worth mentioning that we do not provide any information about intimacy or mentor-mentee relationships during training, ensuring that the graph remains unweighted and homogeneous.

**Online experiments.** For online experiments, we collect a week's data from an in-game event where players invite friends to form teams based on system recommendations. The event unlocks special tasks for team members, offering rewards. We provide an in-game module to recommend an ordered list of friends to each player. When player $u$ accesses the friend recommendation module in GAME, $u$ sees six recommended friends each time. This generates an *exposure* record in the recommendation logs, and $u$ can decide to click on a friend or not. If $u$ is not interested in the current friend list, $u$ can switch to the next recommended result. User $u$ can click on only one friend per day. Once user $u$ clicks on a recommended friend $v$, it sends a team request and generates a *click* record in the recommendation logs. The request requires approval; the clicked friend $v$ can decide to accept or reject it. If $u$ and $v$ successfully form a team, the recommendation module generates a *success* record.

We evaluate DAG, CCGC, and the baseline in the friend recommendation task using three metrics: *(i)* Click Rate, which is the proportion of *click* friends among *exposure* records; *(ii)* Team Rate, the proportion of *success* teams among *click* invitations; and *(iii)* overall Success Rate, the proportion of *success* teams among *exposure* records. The overall Success Rate is the product of the Click Rate and the Team Rate after invitations are sent, *i.e.*, Success Rate = Click Rate × Team Rate. We compare the effects of three strategies:
- Baseline: Rank all friends based on their historical team count.
- DAG: First recall friends in the same community determined by DAG, then rank by historical team formation count.
- CCGC: First recall friends in the same community determined by CCGC, then rank by historical team formation count.

During the week-long event, we train community detection models and output their friend ranking results every day. Each *DAU* is randomly assigned with an algorithm that generates the recommendation results. We finally take the average metrics for one week to ensure a fair comparison. The results, as shown in Table 5, reveal that DAG outperforms the best SOTA method by 7.35%, 1.97%, and 5.24% for the overall success rate, click rate, and team formation success rate, respectively. This superior online performance indicates that DAG can detect more meaningful user communities, *i.e.*, users within the same community have a higher tendency to interact.

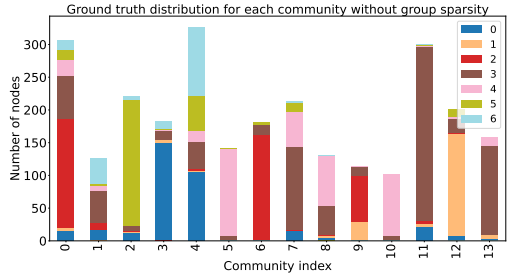

**Figure 8: The ground truth label distribution of detected communities of DAG without group sparsity.**

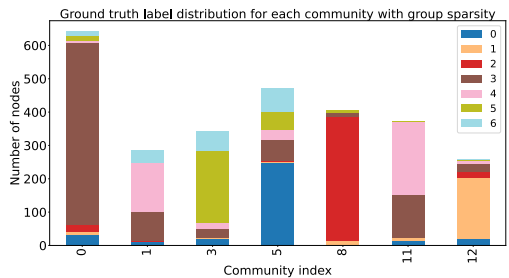

**Figure 9: The ground truth label distribution of detected communities of DAG with group sparsity.**

Furthermore, it is observed that methods with higher EDGE scores also lead to a higher interaction tendency among users. This provides an intuition that the EDGE metric, introduced in this paper, serves as a reliable indicator of the quality of community detection in terms of promoting user interactions. As a result, the EDGE metric not only evaluates the performance of community detection methods but also captures the practical impact of the detected communities on user interactions in real-world scenarios.

In summary, the results of the online experiment demonstrate the effectiveness of the DAG method in detecting meaningful user communities and promoting user interactions. The EDGE metric serves as a robust evaluation measure, highlighting the advantages of the DAG method over existing SOTA in real-world applications, *i.e.*, friend recommendation tasks in online games.

## 6 RELATED WORK

This section briefly reviews related work in traditional community detection methods, deep graph clustering, and masked attribute reconstruction.

**Traditional community detection methods.** Traditional community detection algorithms are based on optimizing modularity and other quality metrics, such as the greedy method [8, 18] and Louvain [4]. Label Propagation Algorithm (LPA) based methods [27, 42] propagates community labels through the graph. These traditional methods initially assume a large maximum number of communities and then merge small communities to optimize unsupervised metrics, finding an adaptive community number $K$ [24]. However, these methods do not take into account the node attributes, which are essential for attributed graphs [37], leading to sub-optimal results and biased community structures. Probabilistic graphical model-based methods, such as SBM [19] and MMSB [1] also can not automatically determine the number of communities.

**Deep graph clustering (DGC).** Deep Graph Clustering (DGC) [13, 16, 24, 34, 41] methods employ Graph Neural Networks (GNNs) and unify the learning from both topological structure and node semantic attributes by learning "clustering-friendly" node embeddings [30, 31, 38], leading to superior performance in community detection tasks. Among these methods, DAEGC [53] uses an attention network [52] and a self-training graph clustering process that jointly optimizes graph embeddings and clustering. CommDGI [61] focuses on community detection with a mutual information mechanism and a clustering layer. AGCN [40] employs fusion modules to fuse node attribute features and topological graph features dynamically. HSAN [35] is a contrastive DGC method that introduces a comprehensive similarity measure criterion and a sample weighing strategy. CCGC [57] mines intrinsic supervision information from high-confidence clustering results and constructs positive and negative samples. However, they need prior knowledge about community number $K$, which precludes real-world application.

**Masked attribute reconstruction.** GraphMAEs [20, 21] employ masked attribute reconstruction to learn node embeddings, achieving the SOTA in downstream node classification tasks.

In this work, we propose DAG to bridge the gap between traditional and deep learning-based community detection methods. Our approach employs a differentiable Community Number Search method, inspired by the traditional community detection methods, to adaptively find the best $K$ during end-to-end training. We also introduce a Masked Attribute Reconstruction module to learn node representations. The proposed method effectively addresses the challenges of $K$-free community detection in attributed graphs.

## 7 CONCLUSION

In this paper, we address $K$-free community detection in attributed graphs by proposing a novel deep learning-based framework, Deep Adaptive and Generative (DAG). DAG detects network communities and searches for the community number $K$ end-to-end without requiring prior $K$. We also introduced the EDGE metric, which is low-cost and robust to varying $K$. Our experiments on public datasets and a real-world social network demonstrated that DAG consistently outperforms SOTA DGC competitors. In conclusion, DAG offers a promising solution for $K$-free community detection, with the EDGE metric serving as a reliable evaluation measure.

## ACKNOWLEDGEMENT

This work is supported by the National Nature Science Foundation of China under Grant 62176155 and Shanghai Municipal Science and Technology Major Project under Grant 2021SHZDZX0102.

| Metric | Modularity | Semantic | NMI | EDGE | K |
|---|---|---|---|---|---|
| Ground Truth | 0.6401 | 11.9360 | N/A | N/A | 7 |
| K-Means | 0.1933 | 20.9624 | 0.1479 | 0.5622 | 7 |
| Greedy Q | 0.8069 | 2.0212 | 0.4673 | 0.8864 | 106 |
| Louvain | 0.8135 | 2.1077 | 0.4468 | 0.8787 | 104 |
| Hanp | 0.6263 | 1.5682 | 0.4010 | 0.7508 | 553 |
| LPA | 0.6605 | 1.5985 | 0.4142 | 0.7871 | 481 |
| DAEGC | 0.7545±0.0011 | 9.5154±0.4359 | 0.4587±0.0285 | 0.8714±0.0120 | 10.0±1.0 |
| CommDGI | 0.6266±0.0300 | 8.9907±2.2156 | 0.3192±0.0193 | 0.8564±0.0234 | 9.4±2.7 |
| AGCN | 0.4338±0.0009 | 20.6211±0.0414 | 0.2172±0.0044 | 0.8297±0.0011 | 10.6±1.2 |
| HSAN | 0.6862±0.0547 | 14.0055±1.6272 | 0.4497±0.0605 | 0.8775±0.0013 | 4.8±1.1 |
| CCGC | 0.6738±0.0690 | 11.8811±1.1215 | 0.5051±0.0546 | 0.8887±0.0141 | 8.5±0.5 |
| DAG | 0.6999±0.0125 | 9.8684±0.7115 | 0.5171±0.0053 | 0.9004±0.0037 | 7.4±1.2 |

Table 6: Cora

| Metric | Modularity | Semantic | NMI | EDGE | K |
|---|---|---|---|---|---|
| Ground Truth | 0.5470 | 11.6463 | N/A | N/A | 6 |
| K-Means | 0.2970 | 19.3495 | 0.2221 | 0.6554 | 6 |
| Greedy Q | 0.8736 | 1.6109 | 0.3378 | 0.8395 | 488 |
| Louvain | 0.8919 | 1.6155 | 0.3243 | 0.8321 | 469 |
| Hanp | 0.6019 | 1.4200 | 0.3402 | 0.7393 | 508 |
| LPA | 0.7177 | 1.6151 | 0.3377 | 0.7530 | 959 |
| DAEGC | 0.7676±0.0052 | 6.8796±1.0481 | 0.2907±0.0070 | 0.8302±0.0036 | 11.5±0.5 |
| CommDGI | 0.7285±0.0041 | 6.5277±0.8006 | 0.2911±0.0041 | 0.8269±0.0018 | 11.1±1.9 |
| AGCN | 0.7624±0.0064 | 7.2502±0.8901 | 0.3160±0.0039 | 0.8316±0.0053 | 8.2±1.1 |
| HSAN | 0.7041±0.0029 | 12.1715±0.0757 | 0.3128±0.0045 | 0.8413±0.0018 | 5.1±0.3 |
| CCGC | 0.7753±0.0021 | 8.4104±0.0852 | 0.4090±0.0050 | 0.8447±0.0037 | 11.9±0.3 |
| DAG | 0.7435±0.0194 | 9.1846±0.0730 | 0.4118±0.0022 | 0.8677±0.0041 | 6.4±0.5 |

Table 7: CiteSeer

| Metric | Modularity | Semantic | NMI | EDGE | K |
|---|---|---|---|---|---|
| Ground Truth | 0.4318 | 200.3377 | N/A | N/A | 3 |
| K-Means | 0.3490 | 435.9176 | 0.3111 | 0.8538 | 3 |
| Greedy Q | 0.7278 | 9.8667 | 0.2217 | 0.8584 | 114 |
| Louvain | 0.7695 | 37.0698 | 0.2062 | 0.8359 | 39 |
| Hanp | 0.3035 | 6.0354 | 0.1770 | 0.7126 | 2037 |
| LPA | 0.6159 | 2.8074 | 0.1804 | 0.7329 | 1924 |
| DAEGC | 0.4989±0.0788 | 170.8658±46.8934 | 0.1784±0.0601 | 0.8422±0.0183 | 4.1±1.1 |
| CommDGI | 0.5562±0.0697 | 161.2432±39.1524 | 0.1892±0.0595 | 0.8469±0.0086 | 4.9±0.7 |
| AGCN | 0.6409±0.0385 | 193.8005±23.4665 | 0.2275±0.0423 | 0.8504±0.0134 | 3.8±0.7 |
| HSAN | OOM | OOM | OOM | OOM | OOM |
| CCGC | 0.5796±0.0712 | 220.9804±58.2542 | 0.1922±0.0035 | 0.8520±0.0172 | 4.1±1.2 |
| DAG | 0.5939±0.0507 | 189.2995±37.5518 | 0.2828±0.0143 | 0.8938±0.0057 | 3.4±0.5 |

Table 8: PubMed

| Metric | Modularity | Semantic | NMI | EDGE | K |
|---|---|---|---|---|---|
| Ground Truth | 0.5420 | 11.3686 | N/A | N/A | 17 |
| K-Means | 0.2061 | 24.9865 | 0.4281 | 0.7793 | 17 |
| Greedy Q | 0.6387 | 2.2243 | 0.4358 | 0.8478 | 90 |
| Louvain | 0.7112 | 3.5303 | 0.4559 | 0.8583 | 64 |
| Hanp | 0.2702 | 2.8806 | 0.4995 | 0.7135 | 885 |
| LPA | 0.6438 | 1.9228 | 0.4858 | 0.8410 | 396 |
| DAEGC | 0.4884±0.0115 | 6.1599±0.3052 | 0.2200±0.0082 | 0.7235±0.0310 | 25.0±2.5 |
| CommDGI | 0.3957±0.0021 | 6.4033±0.4709 | 0.1839±0.0245 | 0.7373±0.0393 | 31.1±1.3 |
| AGCN | 0.4344±0.0021 | 8.7744±1.1372 | 0.1962±0.0187 | 0.7431±0.0105 | 22.6±1.7 |
| HSAN | 0.6259±0.0100 | 7.1363±0.6481 | 0.4131±0.0049 | 0.8375±0.0141 | 29.5±0.5 |
| CCGC | 0.6166±0.0114 | 12.6023±1.0895 | 0.4079±0.0236 | 0.8467±0.0101 | 21.8±3.6 |
| DAG | 0.5981±0.0103 | 14.0695±1.3280 | 0.4320±0.0074 | 0.8629±0.0087 | 15.7±0.9 |

Table 9: Wiki

| Metric | Modularity | Semantic | NMI | EDGE | K |
|---|---|---|---|---|---|
| Ground Truth | 0.5417 | 10.4687 | N/A | N/A | 70 |
| K-Means | 0.2061 | 24.9865 | 0.4281 | 0.7793 | 70 |
| Greedy Q | 0.7270 | 1.9472 | 0.4075 | 0.7290 | 499 |
| Louvain | 0.8126 | 2.3447 | 0.4792 | 0.7300 | 404 |
| Hanp | 0.6670 | 1.9020 | 0.5560 | 0.6670 | 2113 |
| LPA | 0.6466 | 1.8934 | 0.5664 | 0.6705 | 2328 |
| DAEGC | 0.6818±0.0401 | 5.8318±0.3930 | 0.4503±0.0358 | 0.6882±0.0207 | 60.1±0.3 |
| CommDGI | 0.6875±0.0173 | 6.5136±1.2660 | 0.4467±0.0099 | 0.6756±0.0108 | 60.3±0.5 |
| AGCN | 0.6878±0.0147 | 5.7370±0.3423 | 0.4721±0.0108 | 0.6955±0.0065 | 74.2±1.9 |
| HSAN | OOM | OOM | OOM | OOM | OOM |
| CCGC | 0.7176±0.0122 | 6.5657±0.5816 | 0.4898±0.0035 | 0.7047±0.0072 | 73.6±5.4 |
| DAG | 0.6602±0.0101 | 7.5734±0.2918 | 0.4932±0.0037 | 0.7311±0.0076 | 68.4±1.4 |

Table 10: CoraFull

## A PROOF OF GROUP SPARSITY LEMMA

Here, we provide the proof of Lemma 1 (group sparsity).

PROOF. First, in Eq. (11), we add the $\mathcal{L}_{2,1}$ norm constraint on $\mathbf{W}^T$, which makes $\mathbf{W}$ have column sparse characteristics. Because the $\mathcal{L}_2$ norm is used for each column of the $\mathbf{W}$, then the $\mathcal{L}_1$ norm is used for that vector. The $\mathcal{L}_1$ norm is often used to promote sparsity [15], which means that many elements of a vector will be zero. However, since the $\mathcal{L}_2$ norm is calculated first and then the $\mathcal{L}_1$ norm, this will ultimately encourage $\mathbf{W}$ to be column-wise sparse. The $\mathcal{L}_{2,1}$ norm constraint is also a common practice in many other studies [33, 63].

Secondly, the corresponding column of the matrix product $\hat{\mathbf{A}}\mathbf{H}_C\mathbf{W}$ will also be a zero vector like $\mathbf{W}$. In Eq. (16), we represent the matrix $\mathbf{W} \in \mathbb{R}^{d \times k_{\max}}$ as a combination of column vectors and represent any matrix $\mathbf{S} \in \mathbb{R}^{N \times d}$ as a combination of row vectors. Then the result of matrix product $\mathbf{SW} \in \mathbb{R}^{N \times k_{\max}}$ is as Eq. (17).

$$\mathbf{W} = \begin{bmatrix} w_1 & w_2 & \cdots & w_{k_{\max}} \end{bmatrix}, \mathbf{S} = \begin{bmatrix} s_1^T & s_2^T & \cdots & s_N^T \end{bmatrix}^T. \quad (16)$$

$$\mathbf{SW} = \begin{bmatrix} s_1^T w_1 & s_1^T w_2 & \cdots & s_1^T w_{k_{\max}} \\ s_2^T w_1 & s_2^T w_2 & \cdots & s_2^T w_{k_{\max}} \\ \vdots & \vdots & \ddots & \vdots \\ s_N^T w_1 & s_N^T w_2 & \cdots & s_N^T w_{k_{\max}} \end{bmatrix}. \quad (17)$$

If the $j$-th column vector $w_j$ of $\mathbf{W}$ is a zero vector, then elements of $j$-th column of $\mathbf{SW}$ $\{s_i^T w_j, i \in [N]\}$ are all zeros. It can be clearly seen from the vectorized multiplication process that no matter what the value of the left matrix $\mathbf{S}$ is, $\mathbf{SW}$ maintains the same column sparsity property as $\mathbf{W}$. Taking $\mathbf{S} = \hat{\mathbf{A}}\mathbf{H}_C$, we get sparsity of $\hat{\mathbf{A}}\mathbf{H}_C\mathbf{W}$.

Thirdly, for the ReLU function $f(x) = \max(0, x)$, because $f(0) = 0$, the value of the zero element will not be changed.

Following the above three steps, we can conclude that the $\mathbf{C}$ matrix in Eq. (10) has the property of column sparseness. □

## B REPRODUCIBILITY DETAILS

### B.1 Experimental environments.

The proposed DAG and the competitors are implemented with PyTorch [39] (2.0.1) and the DGL (1.1.1+cu117). Each experiment is implemented on an NVIDIA Tesla T4 GPU with 16 GB GPU memory. We train all deep learning-based methods with ten runs and report the average performance. We import modularity from NetworkX (2.8.4), Calinski Harabasz score, and NMI from the scikit-learn library (1.2.2).

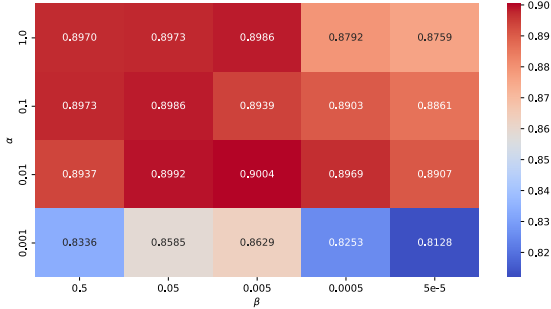

**Figure 10: The Hyper-parameter search for the $\alpha$ and $\beta$ in Cora. The numbers are shown in the heatmap of the average EDGE metric.**

### B.2 Hyper-parameter settings.

All DAG models share the following hyperparameters: GAT layers are used as both the attribute encoder and decoder. The activation function is ReLU. The Adam optimizer is utilized for optimization with a learning rate (lr) of 0.001. The input layer dropout rate is set to 0.2. We set the $\alpha$ to 1e-2 and $\beta$ to 5e-3 in Eq. (13) for all datasets to ensure that our method can both efficiently and effectively adaptive to the real community structures rather than DGC methods' inevitable fine-tuning the community number $K$.

**Dataset-Specific hyper-parameters.** For **Cora**, we the length of embedding (num_hidden) to 512, the number of GAT's attention heads (num_heads) to 4, the number of layers for both attribute encoder and attribute decoder (num_layers) to 2, the weight decay of Adam (weight_decay) to 1e-3, and the maximum number of epochs (max_epoch) to 1500. For **Citeseer**, we set the num_hidden to 256, num_heads to 2, num_layers to 2, weight decay to 1e-4, and the max_epoch to 500. For **Pubmed**, we set the num_hidden to 1024, num_heads to 4, num_layers to 2, weight decay to 1e-2, and the max_epoch to 1000. For **Wiki**, we set the num_hidden to 512, num_heads to 2, num_layers to 2, weight decay to 1e-5, and the max_epoch to 1500. For **CoraFull**, we the num_hidden to 512, num_heads to 4, num_layers to 2, weight decay to 1e-5, and the max_epoch to 1000. The hyper-parameters of **GAME** are the same as CoraFull.

## C IMPACT OF HYPER-PARAMETER.

We adopt a two-step strategy for searching the hyperparameters $\alpha$ and $\beta$ in Eq. (13). Firstly, we conduct a coarse search for $\alpha$ and $\beta$ in a log scale (e.g., [1, 0.1, 0.01, 0.001, 0.0001]). Secondly, we adjust the $\beta$ using a combination of log scale and grid search strategies. For instance, since the Cora dataset's optimal $\beta$ is around 1e-3 to 1e-2, we search values among 2e-3, 3e-3, and up to 9e-3. Finally, we fix the $\alpha$ and $\beta$ for all compared datasets, including the GAME graph. Note that as mentioned in Sec 4, we address the community detection in an **unsupervised manner** and search the $\alpha$ and $\beta$ with the highest product of two unsupervised metrics, *i.e.,* modularity and Calinski Harabasz score. The hyperparameter search results for the $\alpha$ and $\beta$ in the Cora dataset are visualized in Fig. 10 as a heatmap. Based on the heatmap in Fig. 10, we can analyze the impact of varying the hyperparameters $\alpha$ and $\beta$ on the performance of our method, as measured by the EDGE metric.

It can be observed that the best performance is achieved with a value around 1e-3. This suggests that balancing the two hyperparameters yields the most effective community detection results. We can also notice that the performance is relatively stable across different values of $\alpha$ and $\beta$ in longitude scales. Specifically, since the best SOTA method's EDGE metric is 0.8887, there are **60% of cases in the longitude scales outperform the SOTA methods,** indicating that our method is robust to hyper-parameter variations. In conclusion, the hyper-parameter search results for the Cora dataset demonstrate that our method achieves the best performance when a balance between $\alpha$ and $\beta$ is maintained. The robustness of our method to hyperparameter variations further validates its effectiveness in community detection tasks.

## D DETAILED RESULTS ON PUBLIC DATASETS

We provide detailed tables of experimental results for each dataset from Tab.6 to Tab. 10, including modularity, Semantic (Calinski Harabasz score) as unsupervised metrics, as well as NMI and EDGE metrics as supervised metrics. For deep learning-based methods, including SOTA DGC methods and DAG, we provide the mean and standard deviation (*i.e.,* mean ± std) of 10 runs. For traditional methods, we report single-run results. The $K$-means in the tables already use ground truth as prior knowledge. Surprisingly, deep learning-based methods can find more "reasonable" communities than ground truth labels. For example, as shown in Tab. 9, the communities detected by the CCGC and DAG have both higher modularity (tighter internal connections) and higher semantic scores (more attribute similarity) than the ground truth in the Wiki dataset. This shows that in real-life scenarios, there are some hidden factors in the reasons for the generation of communities, which reveals further research directions in community detection algorithms.